\documentclass[epj]{svjour}
\usepackage{graphics}
\usepackage{epsfig}
\begin{document}
\title{Lattice calculations of hadron properties}
\author{Gunnar S.\ Bali\thanks{Invited talk presented at
``Fourth International Conference on 
Perspectives in Hadronic Physics'', ICTP, Trieste,
12 -- 16 May 2003.}}
\institute{Department of Physics and Astronomy, The University of Glasgow,
Glasgow G12 8QQ, Scotland}
\date{August 13, 2003}
\abstract{Recent lattice studies of hadron properties, in particular
of exotic states and charmonia are reviewed. Sea quark and quark mass
effects are discussed as well as decays and mixing.
 % end of PACS codes
} %end of abstract
\maketitle
\section{Introduction}
\label{intro}
While the simplicity and elegance of QCD is very appealing theoretically,
the phenomenological observations of spontaneous chiral symmetry breaking
and even more so of the confinement of colour charges turned it
into a major calculational nightmare: it took almost twenty years after
the discovery of asymptotic freedom to convincingly
demonstrate that the QCD Lagrangian indeed implies these highly
non-trivial collective phenomena. This was done by numerical simulations;
an analytic proof is still lacking.

Fortunately, due to the
property of asymptotic freedom, many short distance/high energy QCD problems 
can be approached by means of perturbation theory. This need not be so
since the very basis of the perturbative expansion
is shaky: confinement implies that quark and gluon fields never appear as
asymptotic states. Fortunately, the success of
jet phenomenology suggests that QCD is reasonably benign in the high
energy region. This is very different in the low energy
regime of {\em strong QCD}. Among the few analytical tools that exist
are the strong coupling and the $1/N$ expansions,
effective field theories (EFTs) and
various QCD inspired or phenomenological models.

QCD can in principle be solved rigorously by means of lattice simulations on a
computer. In practise however computational resources are finite.
This leads to pion masses of typically more than
400~MeV or to simulations within the quenched approximation, where
sea quark effects are neglected.
While these systematic uncertainties reduce with
faster computers,
improved numerical algorithms and theoretical ingenuity, it will always remain
desirable to combine lattice simulations with EFT methods
or, where necessary, with QCD motivated models as
only relatively simple questions can directly be addressed on the lattice.
In many cases problems of phenomenological interest
factorise naturally into a high energy electro-weak part and a
low energy QCD part which can then be evaluated on the lattice,
for instance electromagnetic form factors and
weak decay matrix elements. In many cases QCD problems can also
be factorised
within the framework of EFTs into low and
high energy parts. Lattice simulations
turn out to be invaluable to gain
insight into the dynamics of QCD as many parameters such as the number of
active quark flavours, quark masses, number of colours, temperature
etc.\ are not limited to their phenomenological values but
can be varied.

Several good reviews of lattice calculations exist
in the literature and I refer to them for details on theoretical
aspects, calculational methods and wider phenomenological
implications~\cite{McNeile:2003dy}.
The topics covered in Chris Michael's
review~\cite{Michael:2003ai} ``Exotics''
have a non-vanishing overlap with this article, albeit written
from a (slightly) different perspective.
In my talk I covered the baryon spectrum and structure as well.
Due to the page limit (and lack of new content)
I refer to my recent review~\cite{Bali:2003wj}.
Subjectively selected highlights since then include
two studies of
generalised parton distributions~\cite{Hagler:2003jd,Gockeler:2003jf} and
a determination of the pion form factor~\cite{vanderHeide:2003ip}.
Some progress has been made
in the calculation of electromagnetic $N$ to $\Delta$ transition
form factors~\cite{Alexandrou:2003ea}, a new study of the mass spectrum
of excited nucleons has appeared~\cite{Brommel:2003jm}
and a first step towards consolidation of previous results was
performed~\cite{Dong:2003zf}.
I will not mention the $\Theta^+$, because there are no
lattice results.
Instead I will concentrate on the spectroscopy of charmonia and exotics.

\section{Glueballs and friends}
It was realised as early as
1974~\cite{Wilson:1974sk} that QCD offered
the possibility of
bound states composed only of energy. These
closed-string-/glue-states/boxcitons
became subsequently
known as glueballs~\cite{Robson:1976ff}.
It is hard to imagine anything that demonstrates
confinement more than the
discovery of quarkless massive bound states.
One can also fantasize that non-perturbative physics with particles similar to
glueballs might play a r\^ole
in future theories {\em beyond the Standard Model}. After all QCD is the
only part of the Standard Model with a chance of a mathematically
rigorous definition. So glueballs,
the simplest possible colour-neutral states, are
a challenge that certainly has to be addressed.

The situation is complicated theoretically
as well as experimentally
by the possibility of
mixing with standard quark model states. Even worse QCD does not 
know what is {\em exotic} and what is not. The very meaning of this
term is {\em ``not quark model''} and it is not at all {\em a priori}
clear what exactly we mean by {\em quark model}.
For instance we can ask ourselves
how closely the quark model describes something as profane as the QCD proton.
While introducing the concept of
constituent quarks eliminates the puzzle that
more than 98~\% of its rest mass is generated by spontaneous chiral symmetry
breaking, i.e.\ by the glue,
it is impossible to attribute
the proton's spin and momentum exclusively to quark degrees of freedom.
Gluodynamics plays an even greater r\^ole in properties of the
$\eta'$ and the $\pi$. So if the gluons already leave big footprints
on {\em quark model states} how can we distinguish these from
{\em glue states}? One possibility that immediately springs to mind
would be to look for {\em spin-exotic}
states: when coupling spin and angular momentum within mesons,
only certain combinations of $J^{PC}$
are allowed while for instance $0^{--},0^{+-},1^{-+},2^{+-}$ are forbidden.
Again the situation is complicated by the possibility of $q\bar{q}q\bar{q}$
molecular states to couple to these same quantum numbers.

One toy model is the pure Yang Mills theory of gluodynamics.
In this fictitious world no quark fields exist but
many central features of QCD including confinement and
asymptotic freedom are still reproduced. This well defined
and self consistent theory yields a rich spectrum of
glueballs, all of which are absolutely stable, with the
exception of very heavy ones
that decay into lighter glueballs.

One can go one step further towards QCD by including
quark fields that propagate in the gluodynamic background
but whose feedback onto the vacuum is neglected, the so-called quenched
approximation. Quenched QCD (qQCD)
is no quantum field theory since unitarity is violated, however,
even in this approximation chiral symmetry is spontaneously
broken.
Neglecting the feedback of slowly moving heavy quark sources
onto their environment is a very natural thing to do.
Quenching is also justifiable in
the limit of the number of colours $N\rightarrow\infty$, however,
it is not always clear whether $1/3\ll 1$.
Simulations of pure
$SU(N_c)$ gauge theories
for~\cite{Lucini:2002ku,Lucini2} $N=2, \ldots,6$
seem to indicate this.
As it should be
ratios of light hadron masses from lattice simulations of
qQCD
have been found to be inconsistent with
the observed spectrum~\cite{Aoki:2002fd} however
the differences are typically smaller than 10~\% suggesting
that the quenched approximation has some predictive power
if cautiously consumed.
The consequences of violating unitarity
at light quark mass can become dramatic in some channels
and in particular
in the scalar sector~\cite{Bardeen:2001jm}: roughly speaking as
the axial anomaly does not exist in qQCD
the $\eta'$ will be a surplus light Goldstone boson. The impact
of this can be investigated in quenched chiral perturbation theory
where diagrams that include transitions from scalar to two $\pi$s
yield unwanted contributions that explode as the $\pi$s, including
the would-be $\eta'$, become light.

One justification for quenching is that
the computational effort is easily reduced by a factor of $10^3$.
So it makes complete sense to learn and to understand statistics
and systematics from a quenched study, prior to doing the
real thing. Moreover, most models used in hadron physics
neglect quark pair creation and annihilation, so the model
builders can still learn from qQCD. Also,
``un-quenching'' a model to compare it with lattice results
is easier than un-quenching the lattice simulation.
Last but not least, life is easier in quenched as
lattice studies of strong decays
and excited states are notoriously complicated: glueballs will
not mix with quark model mesons, spin exotic
meson-gluon hybrids are distinct from mesons, are stable and do not 
mix with four-quark molecules either.

\subsection{Heavy glueballs and charmonia}
\label{sec:heavy}
\begin{figure*}
\rotatebox{270}{\epsfig{file=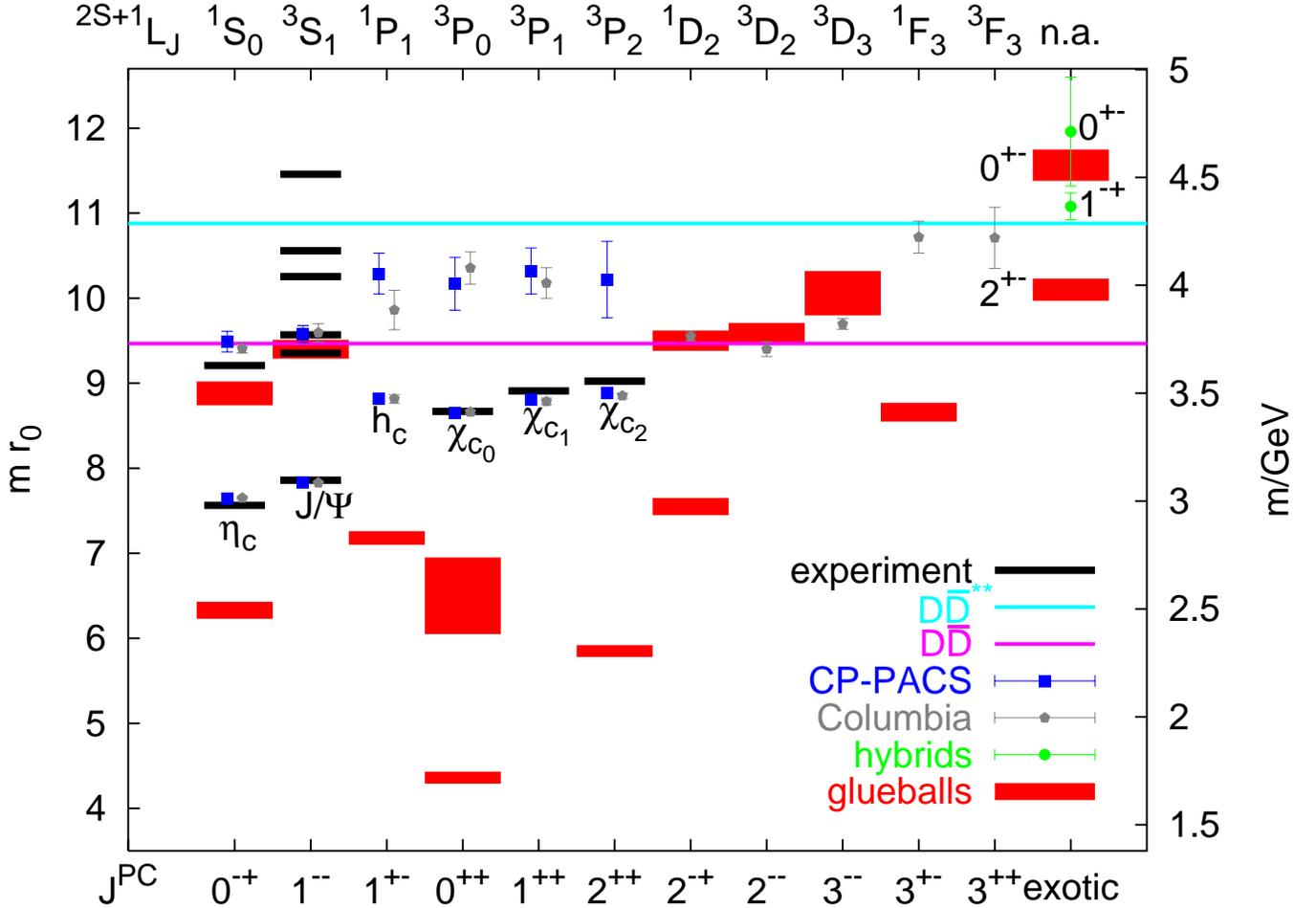,height=.995\textwidth}}
\caption{The quenched charmonium spectrum (CP-PACS~\cite{Okamoto:2001jb},
Columbia~\cite{Chen:2000ej,Liao:2002rj}),
glueballs~\cite{Morningstar:1999rf,Lucini2} and spin-exotic
$c\bar{c}$-glue hybrids~\cite{Liao:2002rj}, overlayed with the
experimental spectrum.}
\label{fig:charm}       % Give a unique label
\end{figure*}

In Fig.~\ref{fig:charm} we compile the glueball spectrum (filled boxes)
of gluodynamics~\cite{Lucini2,Morningstar:1999rf}.
The scale $r_0^{-1}\approx 394$
is set from potential models. Since this is not
the real world a systematic scale error of about 10 \%
should be assumed, for stable states that do not mix!
While these newer results agree with the spectrum of
Ref.~\cite{Bali:1993fb}, during the past decade the statistical
errors have been reduced by factors varying between 1.5 and 2,
depending on the channel. It has also been established that
the $3^{++}$ glueball is lighter than the $1^{++}$. This had
only been suspected in the earlier reference, where the $J$ assignment
of this state remained ambiguous.

The lightest glueball is a scalar, followed by a tensor and a pseudoscalar.
Possibly the lightest state with exotic quantum numbers is a $2^{+-}$ glueball
around 4~GeV. Qualitative features can be understood
in a bag model while the flux tube model seems inadequate~\cite{Juge:2003qd}.
Without numerical simulation it would have been completely impossible
to come close even to a semi-quantitative understanding of these
truly unconventional bound states.

We include experimental charmonium levels (lines) for comparison.
The ${}^{2S+1}L_J$ notation refers to these states. We have only
included confirmed resonances, with masses taken from the
Particle Data Book. For the $\eta_c'$ we took the
Belle result from double charmonium production~\cite{Abe:2003ja},
3630(8)~MeV. One often thinks of the charm quark
as heavy in the sense that its mass is much bigger than mesonic and
baryonic QCD binding energies.
However, the spectrum of states entirely made out of glue with no quarks
at all covers a similar energy range!
We also display recent quenched lattice results
from two groups~\cite{Okamoto:2001jb,Chen:2000ej,Liao:2002rj} (CP-PACS and
Columbia). These have been obtained using relativistic charm quarks
on anisotropic lattices
with the ratio of spatial over temporal lattice
spacings $\xi\approx 2$ (Columbia)
and $\xi\approx 3$ (CP-PACS).
A recent simulation of $J\leq 1$ $S$ and $P$ wave charmonia on isotropic
($\xi=1$) lattices by
QCD-TARO~\cite{Choe:2003wx}
(who also study charmonium wave functions)
confirms these findings. Note that in all these simulations the effect of
diagrams with disconnected quark lines has been neglected. One might expect
OZI violating
contributions from these, in particular for states that lie close
to glueballs with the same quantum numbers.

The charm quark mass has been adjusted such that the spin averaged $1S$
state reproduces experiment. This means that in this case
the quenching errors
of up to 10~\% apply to
spin-averaged level splittings with respect to this ground state,
rather than with respect to zero energy.
This systematic scale uncertainty caused by omitting sea quarks
renders it irrelevant whether we use a ``constructed'' scale like $r_0$
or an experimental mass like the $1P-1S$ gap  as an input (which would
increase the splittings by about 6~\%).
For the fine structure, potential models combined with
lattice results~\cite{Bali:1998pi} tell us that we should expect
to undershoot the real world
number by up to 40 \%. For states above threshold or where mixing effects
and strong decays play a big r\^ole radical
changes might occur while for the $2P$ excitations finite
volume effects could be an issue.

There are different ways of obtaining a given $J^{PC}$:
for instance $J^{PC}=1^{--}$ can be
an $S$ or a $D$ wave.
The correlation function associated with the 
respective $D$ wave operator decays very fast in Euclidean time, into
the same ground state as the respective $S$ wave: it appears
that the charm quark is
too light to turn $L$ into an (approximately) good quantum number.
A similar behaviour has been observed for the
$2^{++}$ and $1^{++}$ states (either $P$ or $F$ waves).

In addition to the standard charmonia and glueball states the
figure contains the lightest two
spin exotic $c\bar{c}$-gluon
hybrids~\cite{Liao:2002rj}.
At least two other studies of
hybrid charmonia with relativistic charm quarks
exist to-date~\cite{Mei:2002ip,Bernard}. Given the fact that not even $L$ is
a good quantum number it is not clear how one would distinguish
non spin-exotic hybrids from conventional radial excitations.
The lightest exotic hybrid turns out to be a vector, $1^{-+}$. This
is followed by $0^{+-}$ and $2^{+-}$ hybrids.

Once sea
quarks are switched on, most of the states calculated on the lattice
will decay strongly and mixing will
occur too. In this context it is interesting to see that
the spin-exotic $0^{+-}$ glueball and $c\bar{c}g$ hybrid
have similar masses.
We have included two experimental thresholds into the figure:
$1^{--}$ charmonia can and will decay into a $D\overline{D}$ meson pair.
Nonetheless resonances that exceed this threshold
by almost 1~GeV have turned out to be experimentally
detectable. The exotic $1^{-+}$ hybrid cannot decay into
an identical boson-antiboson pair since $PC=-$. The next possibility,
a $\overline{D}$ and a vector $D^*$, is suppressed both in flux tube
models~\cite{Page:1998gz}
where the quark-antiquark pair that breaks
the flux tube is produced in its centre,
as well as in the heavy quark limit in which the
angular momentum of the heavy quarks is fixed. The next possibility,
which follows this $S+P$ selection rule, would be
a decay into a $\overline{D}$ and a $P$ wave isovector
$D^{**}[=D_1(2420)]$.
Interestingly the mass of the $1^{-+}$ is compatible with this threshold
such that mixing with (would be) molecular states is an issue. Should the
state be below this threshold a resonance with a width smaller than
100~MeV might be possible, with dominant decay into
a $\chi_c$ under emission of a light scalar meson~\cite{McNeile:2002az}.

At present we do not know whether the overlap with glueball states
in the quenched approximation has any implications on the phenomenology
of charmonia. Na\"{\i}vely the phase space for decay of heavy glueballs
appears huge with no quarks to preserve and more than 3 GeV of energy to
disperse. However, before drawing definite conclusions dynamical issues
related to the glueball wave functions need to be investigated.
In particular mixing is a strong possibility:
for instance the $\psi(3770)$,
$\psi(2S)$, $D\overline{D}$ threshold and the vector glueball
all lie within the same 100~MeV mass window, allowing for a
potentially interesting
phenomenology, in particular since there is little evidence from
lattice simulations to support distinguishable $2S$ and $1D$ levels
in this region.
Recently, the interesting question
of a (non-exotic) hybrid component in the $1S$ states has been addressed
in a very nice study~\cite{Burch:2003zf}, in the
framework of NRQCD. The conclusion is that this contribution is weak.
The question becomes more exciting but also tremendously less approachable
if the $2S$ charmonium state and
glueball channels were considered in addition.

With data from Belle and Babar emerging, CLEO-c being online
and the possibility of BES III starting to take data in 2006
the charmonium region is also exciting experimentally.
Unfortunately, there is no way of producing say spin-exotic
$1^{-+}$ states at any detectable rate in a decay starting from 
a vector resonance. For this we might have to wait for the
proton-antiproton PANDA experiment at GSI. However, in the meantime
there will definitely be progress in measuring spectrum and
decay rates. Possibly the glue-richness of this mass region
will leave its imprint, for instance by enhancing OZI suppressed
processes.

\subsection{Light scalars: today}
A lot of experimental attention has been devoted to
the spectroscopy of light scalar mesons. The reasons
are threefold: the scalar sector is intimately linked to
chiral symmetry breaking which results in both light $\pi$s
but at the same time also in heavy $\sigma$s. Furthermore, the lightest
glueball is predicted to be a scalar and, as an added extra,
the lightest four-quark candidate state has scalar quantum numbers
as well: pseudoscalar-pseudoscalar bound states can be lighter
then quark model scalars since chiral symmetry breaking
makes the latter ``artificially''
heavy and the former ``artificially'' light.
It is also not completely accidental that gluons play a dominant
r\^ole in both the pseudoscalar and the scalar sectors, in the former
due to the axial anomaly and in the latter due to glueballs, both
of which result in large OZI violations and splittings between
$I=0$ and $I\neq 0$  mesons.
Unfortunately, it is exactly this rich phenomenology
which complicates the interpretation of experimental results as
well as theoretical calculations.

The lightest scalar ``particle'' is
the $\sigma$ [or $f_0(600)$]. It seems by now clear that
this pole shifts the phase of the 
$\pi\pi$ $S$ wave, both from $\pi\pi$ scattering experiments
as well as from proton-antiproton
collisions with a 3$\pi$ final state. With an imaginary part
of the $T$ matrix pole of 450(150) MeV, which almost exceedes
its real part [800(400) MeV], few people are left who
would call this sort of ``resonance'' a ``particle''. However
historically, before the advent of
lattice simulations, the $\sigma$ was among the first glueball candidates.
Next there is the $f_0(980)$ which in contrast has a width not much
larger than 50~MeV, and is accompanied by the $I=1$ $a_0(980)$ triplet.
This ``narrowness'' is due to the fact that these states cannot decay
into $K\overline{K}$, being
10--15 MeV lighter than this threshold. This proximity
also turns them into
natural candidates for either $s\bar{l}l\bar{s}$
or $K\overline{K}$ bound states. For simplicity
We will not distinguish between four-quark and meson-meson bound states
but refer to both as ``molecules''.

Lattice
results~\cite{Bali:1993fb,Sexton:1995kd,Morningstar:1999rf,Lucini2}
suggest that the lightest glueball should be a scalar with mass somewhere
between 1.4 and 1.8~GeV. While all raw lattice data agree within statistical
errors of some 40~MeV, the large uncertainty quoted above is due to the
scale uncertainty
{}from using the quenched approximation.
These results lie on top of three experimental scalar $I=0$ resonances,
the broad  $f_0(1370)$,
the extremely well studied $f_0(1500)$ and the $f_0(1710)$.
The standard picture is that these states are mixtures between
a glueball and two $I=0$ nonet mesons, one with dominantly
$u\bar{u}+d\bar{d}$ and the other with $s\bar{s}$ quark content.
The seven  remaining $I\neq 0$ nonet members are most likely
the four $K_0^*(1430)$ as well as
three $a_0(1450)$ states.
Whether there is an excess of experimental states over quark model
states or not critically depends on how resonances
are organised into nonets.
It has to be said that the $a_0(1450)$ is quite broad and that
the experimental evidence for these states
is not rock hard. On the other hand, with a splitting of 450~MeV,
it is extremely hard to reconcile the $K_0^*$ and the $a_0(980)$ into
the same nonet, in particular
the near degeneracy of the
$a_0$ with the $f_0(980)$ suggests very tiny OZI violating effects.
[This in turn also means that there is little chance for
a significant glueball component in the $f_0(980)$.]
The remaining puzzle is the absence of
$S\neq 0$ partners of the $a_0/f_0(980)$. Most likely these
states have a large molecular
component and are part of an inverted nonet~\cite{Jaffe:1976ig},
together with the $f_0(980)$, a would-be-$\pi \pi$ $\sigma$ state
and would-be-$\pi K$ $\kappa$s,
which just happen to be extremely fragile, due to the light $\pi$
involved.

\subsection{Light scalars: future}
To support the above picture an unambiguous experimental
identification of the $a_0(1450)$ and a theoretical clarification
of the molecular nature of the $a_0/f_0(980)$ states rank high on the
wish list.
Different experimental and theoretical inputs have led to various
mixing models~\cite{Amsler:1995td,Burakovsky:1998zg,Lee:1999kv,Close:2001ga}.
Amsler and Close~\cite{Amsler:1995td} suggest that the 
light quark state mainly goes into the $f_0(1370)$, with a subleading
component mixing into the $f_0(1710)$ while the
$s\bar{s}$ and the glueball are mainly distributed between the
$f_0(1500)$ and $f_0(1710)$. This view has been superceeded by
Close and Kirk's more recent analysis of decays and production
rates~\cite{Close:2001ga} which suggests the
$f_0(1500)$ to be dominantly glueball in character
with light and strange quark
components mixing destructively and additively into
the $f_0(1370)$ and $f_0(1710)$, respectively.
In contrast Refs.~\cite{Burakovsky:1998zg,Lee:1999kv}
predict the $f_0(1710)$ to be the (dominant) glueball-state with
the two lighter resonances being
composed primarily of the quark model mesons.

All these predictions crucially depend on the input parameters used,
in particular on the ordering of ``unmixed'' states and on assumptions
on the mixing matrix. One might expect the unmixed
$d\bar{d}+u\bar{u}$ state to have a mass close to that of the $a_0(1450)$
but unfortunately the latter resonance is very broad experimentally and not
overly well established. While in QCD with light sea quarks all states
will automatically be ``mixed'', the quenched approximation is ideal
to address the question of ordering of ``unmixed'' states and even
for estimating off-diagonal mass-matrix elements.
The main problem however is that in this case
correlation functions will loose positivity at small quark masses
as the $\pi$ and the $\eta'$ are degenerate~\cite{Bardeen:2001jm}.
In practise this means
that at best $s\bar{s}$ scalars can be approached using traditional methods.

In Ref.~\cite{Lee:1999kv} significant finite size effects and a strong
lattice spacing dependence are observed. After extrapolations
the quenched $s\bar{s}$ scalar appears to be about 300~MeV lighter than
the pure glueball.
RBC~\cite{Prelovsek:2002qs} quote an even lower mass of
1.04(7)~GeV for the light quark $a_0$ and 0.9(1)~GeV for the
$I=0$ $f_0/\sigma$. The leading chiral correction from
the $\pi\eta'$ loop was included into this study.
One would however hope that systematic uncertainties
will be identified and investigated in the future.
Should this result be confirmed, then
the $a_0(980)$ could be explained as a quark model state but
what is the $f_0(980)$ and why are the $K_0^*$s,
that are experimentally well established, so heavy?
In a very sophisticated
study Bardeen {\em et al.}~\cite{Bardeen:2003qz,Bardeen:2001jm}
determine the relevant low energy parameters of the quenched chiral Lagrangian
and remove the leading order quenched chiral power and log.
After testing the approach in the pseudoscalar sector
they are able to determine an $a_0$ mass of
1.33(5)~GeV, expressed in units of the $\rho$ mass, which is 
by about
$200\pm 100$~MeV lighter
than the glueball, extrapolated to the continuum limit and
converted into the same units.
The unmixed $s\bar{s}$ state
might then be expected to be fairly equal in mass to the glueball.
The main difficulty is that with only two lattice spacings at hand
the extrapolation to the continuum limit is still a bit shaky.

The situation has also been studied by UKQCD~\cite{Hart:2001fp} in QCD with sea quarks
at lattice spacings
$a\approx 0.1$~fm and $a\approx 0.13$~fm. In this case the
$a_0$ is significantly lighter but the same holds true for
the flavour singlet meson/glueball (see Fig.~\ref{fig:glue} below).

In addition to the studies of quark model scalars some results
exist on four quark molecules, but with degenerate masses,
with the aim of addressing the possibility of a
$\pi\pi$ bound state. Results of quenched studies~\cite{Alford:2000mm}
are inconclusive due to the requirement of large volumes to
avoid squeezing the $\pi$s on top of each other when they intend to
be elsewhere: without a careful finite size study a residual interaction
energy can hardly be disentangled from the $\pi\pi$ scattering
phase~\cite{Luscher:1985dn}.
Diagrams with disconnected quark lines have also been neglected
(with one exception~\cite{Fukugita:1994ve}). In the absence
of definite results in the quenched approximation,
the SCALAR collaboration
has started the very ambitious program of determining the mass of
an $I=0$ $\pi\pi$ state incorporating sea quarks~\cite{Kunihiro:2002pu}.
In view of the above it is highly attractive to study simplified cases,
with heavier quarks and $I=1$, like $K\overline{K}$ or $DK$ 
(see also Sec.~\ref{sec:ds} below) molecules,
both with at least equally important phenomenological implications.

Few exploratory studies of strong decays exist.
Since we are working in Euclidean time
there is no concept of asymptotic states. Neither can we
calculate the imaginary part
of a forward amplitude but there are bug-fixes:
the method of choice is the computationally very challenging
exploitation of finite size effects introduced by
L\"uscher~\cite{Luscher:1985dn}.
The poor man's approximation has been developed by
Gottlieb {\em et al.}~\cite{Gottlieb:1983rh} and Michael~\cite{Michael:1989mf}.
In the latter case an on-shell transition matrix element is evaluated
by adjusting the rest mass of the decaying particle to the energy
of the outgoing state. In QCD with sea quarks this method can
only be approximate as the operator used to create
the initial state will in general also
have a non-vanishing overlap with the final state.
The transition matrix element is related to a coupling which in turn,
assuming momentum independence, can be normalised to phase space,
predicting the partial decay width.
The on-shell condition implies that for a given
mass of the final state
the relative momentum has to be adjusted to guarantee
energy conservation. As momentum is discretized on a lattice this
imposes some constraints.
The initial state can be boosted such that at a fixed mass
several relative momenta can be (approximately) realised, on sufficiently
large volumes, and the momentum
dependence of the coupling checked.
Chiral perturbation theory can also help to connect
results obtained at different quark mass, once these are sufficiently
light.

Promising results using this approach have been obtained
recently for the $\rho\rightarrow\pi\pi$ decay in $n_f=2$ QCD
by McNeile and Michael~\cite{McNeile:2002fh}.
Decay couplings of a glueball to two pseudoscalars
have so far only been computed by GF11~\cite{Sexton:1995kd} some 9 years
ago: a mass dependence has been observed with
a stronger coupling to heavier mesons, which then has to be
folded with phase space.

Elements of the glueball-scalar mass matrix
have been calculated by two groups, again GF11
in qQCD~\cite{Lee:1999kv} and
UKQCD for $n_f=2$~\cite{McNeile:2001xx}, for quark masses
around the strange quark. Both collaborations
obtain
mixing energies of about 300~MeV and 500~MeV on relatively coarse lattices,
respectively.
Starting from two degenerate unmixed states this would imply
level splittings between the mixed states of 0.5 -- 1 GeV.
An extrapolation to the continuum limit by
GF11 who also simulated three finer lattice spacings
resulted in 61(45)~MeV, a very reasonable value but with a
100~\% uncertainty.

\begin{figure}
\resizebox{0.488\textwidth}{!}{%
  \includegraphics{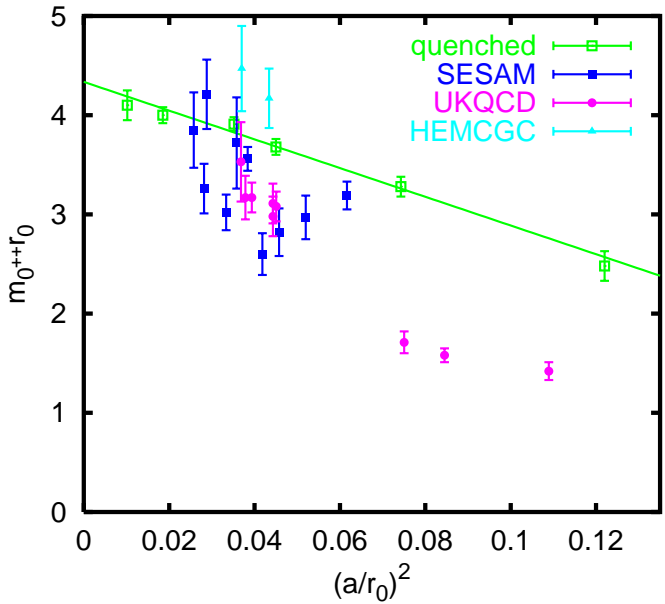}
}
\caption{The scalar ``glueball'': qQCD vs.\ $n_f=2$.}
\label{fig:glue}
\end{figure}
We are still in the position that
the combined ``world data'' on the scalar $n_f=2$ ``glueball'',
plotted as a function of the squared lattice spacing $a^2$,
fits into Fig.~\ref{fig:glue}.
The quenched
case~\cite{Lucini2,Bali:1993fb} is included for reference.
The un-quenched results have been obtained by use of three
different lattice discretizations of the Dirac action:
staggered (HEMCGC~\cite{Bitar:1991wr}),
Wilson (SESAM~\cite{Bali:2000vr})
and clover
(UKQCD~\cite{McNeile:2001xx,Hart:2001fp}).
The quarks are all not much lighter than the strange quark,
the scalar meson is still stable and
the wave function
turns out to be very close to that of the quenched
glueball~\cite{Bali:2000vr,McNeile:2001xx}. Most $n_f=2$
points clearly lie below the quenched line, however, there
is certainly a slope in the results,
such that the mass in the physical $a=0$ limit does not contradict
the quenched result. Within the SESAM data set
there is an apparent discontinuity because different points
have been obtained at different quark masses;
the ``glueball'' becomes lighter
as the quark mass is reduced.
Clearly additional studies at lighter quark masses and different
lattice spacings are required.

\subsection{The $D_{sJ}^+(2317)$}
\label{sec:ds}
Recently a narrow $D_{sJ}^+(2317)$ state has been detected by
BaBar~\cite{Aubert:2003fg}, dominantly decaying
into $D_s\pi$. This finding was
confirmed by CLEO~\cite{Besson:2003jp} and by Belle~\cite{Abe:2003jk}.
The latter two collaborations also reported a narrow resonance around
2537~MeV, decaying into
$D_s^*+\pi$. Both states lie by about 40~MeV below the
respective $DK$ and $D^*K$ thresholds. In contrast potential
model calculations
suggest~\cite{Godfrey:xj} much heavier masses for the
missing $D_{s0}$ and $D_{s1}'$ $P$ wave states, rendering these into
broad resonances. These expectations have very recently
received experimental support from Belle's observation of
the missing $D_0^*$ and $D_1'$ states at $2308 (17)(15)(28)$~MeV and
at 2427 (26)(20)(15)~MeV, respectively~\cite{Abe:2003zm}.
These states indeed
strongly decay into $D\pi$ and $D^*\pi$, respectively, with widths
of order 300~MeV.

If the new $D_{sJ}$ mesons were $P$ wave $c\bar{s}$ mesons, why are
they so light but the corresponding $c\bar{d}$ mesons are not?
Naturally this question invites speculation
that the new scalar state might be of a $DK$ (or 4-quark)
molecular nature, somewhat resembling the $f_0/a_0(980)$
system~\cite{Barnes:2003dj}.
Of course such ideas eventually have to be substantiated by
a QCD calculation and indeed lattice results exist:
simulations in the static limit within the quenched approximation~\cite{Michael:1998sg} and with sea quarks~\cite{Bali:2003jv}, simulations
including NRQCD/HQET $1/m$ corrections~\cite{Hein:2000qu}
and simulations with a relativistic charm
quark~\cite{Boyle:1997rk,Dougall:2003hv}. The interpretation of these
results is controversial: I~\cite{Bali:2003jv}
observed that both effects, including sea quarks and including relativistic
corrections to the static limit,
increase the mass of a quark model $c\bar{s}$ $P$ wave scalar state,
pushing it above the $DK$ threshold and into agreement with potential
model predictions. I then concluded that lattice results
are incompatible with a pure quark model nature of the new $D_{sJ}$ states.
On the other hand, while confirming an increase in the predicted mass when
incorporating sea quarks, the authors of Ref.~\cite{Dougall:2003hv}
conclude that their results are consistent with the states observed
by Babar and CLEO, however, within errors no disagreement is seen
with the potential model predictions either.
Interestingly, while $1/m$ corrections to the static limit
are substantial for charm quarks and increase the
$0^+-0^-$ splitting by about 25~\%~\cite{Bali:2003jv},
the $1^+-1^-$ splitting is found to agree with the $0^+-0^-$
splitting within statistical
and systematic errors of about 15~\%, in quenched as well as
with sea quarks~\cite{Dougall:2003hv}, as suggested by chiral
symmetry in the heavy quark limit~\cite{Nowak:1992um}.

The same chiral symmetry argument would also
apply to molecular states, such that the mere discovery of
similar  $B_{sJ}$  mesons would not help in discriminating
between meson and molecule. Dynamical issues need to be addressed
and decays investigated. In particular electromagnetic decays,
where the theoretical understanding is much better than for strong decays,
would reveal information about the internal structure of these states.
The narrowness of these resonances suggests that this need not
be a completely
hopeless enterprise. If the lightest $D_{sJ}$ state was dominantly
a molecule, then a quark model $J=0$ resonance should exist in
addition. The recent Belle~\cite{Abe:2003zm} evidence of
the $D_0^*(2308)$ suggests that another broad resonance,
centred around 2.4 to 2.5 GeV with dominant decay into
$DK$, might be detectable.

Better lattice studies are needed to clarify these important issues,
and also with respect to the possibility of similar states in the
$B$ meson system. Of course at physical light and strange quark masses
lattice QCD should reproduce the experimental spectrum. So to gain
insight into the nature of these states a high precision
quenched benchmark study is required: in this approximation molecules
and mesons are clearly distinct. In addition,
the light sea quark mass dependence of both, a possible $c\bar{l}l\bar{s}$
molecule and a $c\bar{s}$ meson has to be traced down into the region
where mixing can set in. Fortunately, the $DK$ system is more user
friendly than its $K\overline{K}$ counterpart: heavier particles
can be accommodated in smaller volumes and the binding energy
that is suggested by phenomenology is comfortably large with
about 40~MeV, rather than
a mere 10--15~MeV. It is also conceivable to calculate matrix elements that
are related to electromagnetic decay rates.

\subsection{More hybrids}
\begin{figure}
\resizebox{0.488\textwidth}{!}{%
  \includegraphics{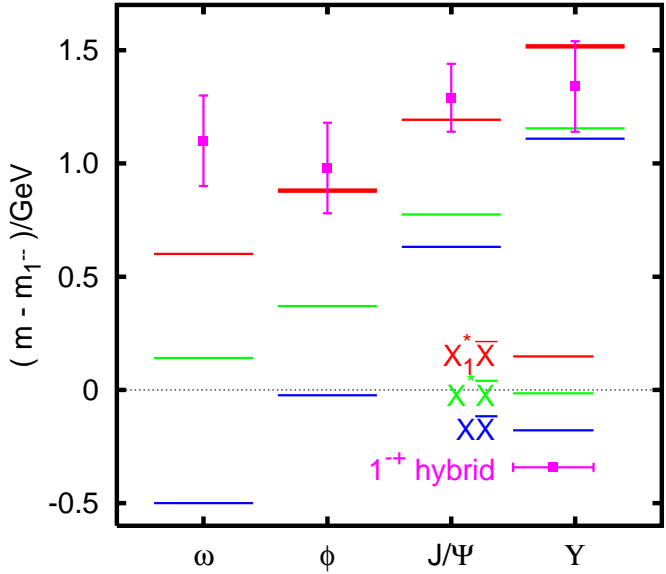}
}
\caption{Splitting of hybrid meson masses with respect to the respective
triplet $S$
wave ground states for the light, strange, charm and bottom cases.
$X$ denotes $\pi,K,D$ and $B$ mesons.
$X^*$ denotes the $\rho,K^*,D^*$ and
$B^*$ vector mesons while 
$X_1^*$ stands for
the (likely) $1^+$ states $b_1(1235)$, $K_1(1270)$, $D_1(2400)$ and
$B^*_J(5732)$, respectively.}
\label{fig:hybrid}
\end{figure}
The spectrum of $c\bar{c}$ glue hybrids has already been discussed in
Sec.~\ref{sec:heavy} above. Light hybrids have been studied to
some extent as well,
with~\cite{Lacock:1998be,Bernard:2003jd} and
without~\cite{Lacock:1996ny,Mei:2002ip,Bernard} sea quarks.
All these results yield the same ordering as in the charmonium case
(and in fact also the bottomonium case)
with $1^{-+}$ being the lightest exotic, followed
by $0^{+-}$ and $2^{+-}$. The $1^{-+}$ is consistent
with a mass of $1.9(2)$~MeV, with the corresponding strange quark exotic
about 200~MeV heavier, in quenched as well as in un-quenched
simulations. This is heavier than the $\pi_1(1600)$ candidate.
Mixing with molecular states is a possible explanation and the
feasibility of studying this has been demonstrated by
MILC~\cite{Bernard}.
However, at present the quarks are still so heavy that 
for instance the combined mass of a $b_1$ and a $\pi$ is around 1.9~GeV
too. Clearly the lattice calculations are incomplete. Molecules
have to be included and the quark mass dependence of the mixing matrix
has to be studied carefully.

Heavy hybrids can be studied using NRQCD~\cite{Collins:1997cb,Manke:1998yg,Manke:1998qc,Juge:1999ie,Manke:2001ft,Burch:2003zf,Juge:2003qd} or
the Born-Oppenheimer approximation~\cite{Perantonis:1990dy,Collins:1997cb,Juge:1999ie,McNeile:2002az,Juge:2003qd}.
In the charmonium case simulations with relativistic quarks
on anisotropic lattices have been pursued
by several groups~\cite{Liao:2002rj,Choe:2003wx,Mei:2002ip}
as well as with isotropic lattices~\cite{Bernard,Bernard:2003jd}.
Even a simulation of bottomonium with relativistic quarks
on an anisotropic lattice exists~\cite{Liao:2001yh}.
With the statistical errors of present simulations differences
between quenched and un-quenched data cannot clearly be resolved.
In Figure~\ref{fig:hybrid} We summarise the present estimates of
the splitting of the $1^{-+}$ hybrid with respect to the respective
vector meson state: the flavour dependence is tiny.
In addition three decay thresholds are displayed. As detailed
in Sec.~\ref{sec:heavy} a strong decay into two pseudoscalars is forbidden
while the decay into vector and pseudoscalar is suppressed, in particular
in the heavy quark limit. Theoretically the $b\bar{b}$ hybrid is most clean-cut
but experimentally hard to produce.

On the lattice electromagnetic matrix elements can be
calculated~\cite{Choe:2003wx,Burch:2003zf} but strong decays are very
challenging. To this end McNeile {\em et al.}~\cite{McNeile:2002az}
predict that
if the lightest bottomonium hybrid is indeed below the
$B^{**}\overline{B}$ threshold the dominant decay channel should be 
deexcitation by emission
of a scalar: $H_b\rightarrow \chi_b\pi\pi$, with a width of about
100~MeV. The same argument should also be valid in the charmonium case,
where phase space would reduce the width even further, but not necessarily
for light hybrids.

\section{Conclusions: from fiction to fact}
A combination of new theoretical methods and computing technology
has allowed us to arrive at the boundary between qualitative
test of principle and quantitative prediction in the complicated area
of flavour singlet physics, strong decays and mixing.
A few years ago at least three major technical challenges had to
be overcome:
light quarks, sea quarks and disconnected quark lines.
We are about half way through by now.
With enough effort devoted onto the topics covered in this article,
quantitative predictions are possible within the time
scale that is relevant for experiments like glueX at JLAB or PANDA at GSI.
\section*{Acknowledgments}
I warmly thank 
Sigfrido Boffi, Claudio Ciofi degli Atti and Mauro Giannini for organizing
this stimulating meeting.
This work has been supported by
PPARC grants PPA/A/S/2000/00271 and PPA/G/0/2002/0463.

\end{document}